
\input phyzzx
\nopagenumbers
\overfullrule0pt
\def\eps{\epsilon}
\def\wtil{\widetilde}
\def\cosbpa{\cos(\beta+\alpha)}
\def\sinbpa{\sin(\beta+\alpha)}
\def\rta{\rightarrow}
\def\mz{m_Z}
\def\mw{m_W}
\def\lam{\lambda}
\def\mhp{m_{H^\pm}}
\def\hp{H^+}
\def\hm{H^-}
\def\wm{W^-}
\def\npbj#1{{\it Nucl. Phys.} {\bf B{#1}}}
\def\crr{\crcr\noalign{\vskip .15in}}
\def\SCIPP{\centerline {\it Santa Cruz Institute for Particle Physics}
  \centerline{\it University of California, Santa Cruz, CA 95064}}
\def\DAVIS{\centerline {\it Department of Physics}
  \centerline{\it University of California, Davis, CA 95616}}

\footline{\hfil\folio\hfil}
\headline{\hfil\null\hfil}
\footline{\hfil\null\hfil}
\Pubnum={$\caps UCD-92-31$\cr$\caps SCIPP-92/59$\cr}
\date={December 1992}
\pubtype{}
\titlepage
\baselineskip 0pt
\title{Errata for {\it Higgs Bosons in Supersymmetric Models: I, II and III}.}
\author{J.F. Gunion}
\vskip .1in
\DAVIS
\author{H.E. Haber}
\vskip .1in
\SCIPP

\abstract
\baselineskip 0pt

Errata are given for J.F. Gunion and H.E. Haber, \npbj{272} (1986) 1,
\npbj{278} (1986) 449, and \npbj{307} (1988) 445.

\endpage
\titlestyle{Errata for J.F. Gunion and H.E. Haber,}
\titlestyle{\npbj{272} (1986) 1.}
\medskip
\pointbegin
At the beginning of section 2, we state that
eq.~(2.1) is the most general two-Higgs doublet
scalar potential subject to a discrete symmetry $\phi_1\rta -\phi_1$
which is only softly violated by dimension-two terms.  This is not
strictly correct.  First, rename $\lambda_7$ appearing in eq.~(2.1)
as $\lambda_8$.  Then, there is one additional term that can be
added:
$$\lambda_7\left[{\rm Re}~\phi_1^\dagger\phi_2-v_1 v_2\cos\xi\right]
\left[{\rm Im}~\phi_1^\dagger\phi_2-v_1 v_2\sin\xi\right]\,.$$
However, this term can be eliminated by redefining
the phases of the scalar fields.  To see this, note that if
$\lambda_7\neq 0$ then the coefficient multiplying
the term $(\phi_1^\dagger
\phi_2)^2$ in the scalar potential is complex, while if $\lambda_7=0$
then the corresponding coefficient is real.  Subsequent results
presented in the paper are not affected by this choice.  Moreover,
in the minimal supersymmetric model, $\lambda_7=0$ (at tree-level).
On the other
hand, in CP-violating two-Higgs doublet models, it is important to
keep $\lambda_7\neq 0$ if one wishes to retain the overall freedom
to redefine the Higgs field phases.
\point
In eq.~(3.3), the sign of the $\mu$ term should be switched.  That is,
the relevant term in the superpotential should read: $W=-\mu\epsilon_{ij}
H_1^iH_2^j$, in the convention that $\epsilon_{12}=-\epsilon_{21}=1$.
Then, the signs of the $\mu$ terms in eq.~(4.28) and in the equations
of Appendix A are all correct.  However, the sign of $\mu$
in eqs.~(4.13) and (4.19) and in the associated squark-squark-Higgs
Feynman rules of Figs. 11, 12, 13, and 14 must be changed.
\point
In Fig. 1, to get the Feynman rules for incoming $\wm$ and outgoing
$\hm$, but leaving the momentum definitions the same,
multiply rules (a) and (b) by $-1$; rule (c) is unchanged.
\point
In eq.~(4.10), a $d$ should appear immediately to the right of
the last square bracket.
\point
In eq.~(4.11), in the term proportional to $m_Z$ replace
$H_1^0\cosbpa+H_2^0\sinbpa$ by $H_1^0\cosbpa-H_2^0\sinbpa$.
\point
In Fig. 9, multiply the $H_1^0H_1^0H_1^0$ and $H_2^0H_2^0H_2^0$
vertices by a factor of $g$.  Otherwise, the Feynman rules of Figs.
9 and 10 are correct as depicted.
\point
A point of clarification.
The reader should note that eqs.~(4.12)--(4.15) contain only those
terms needed to compute the mass and scalar interactions
of the squarks and sleptons.  To obtain the entire scalar potential,
one must add the terms of eqs.~(3.8) and (3.9) that are not
included in eqs.~(4.13)--(4.15).
\point
In Fig. 17(c) and (d),
the Feynman rules for the $H_1^0H_2^0\widetilde q_{kL}
\widetilde q_{kL}$ and $H_1^0H_2^0 \widetilde q_{kR}
\widetilde q_{kR}$ vertices are incorrect.  The correct Feynman
rules are:
$$\eqalign{\qquad H_1^0H_2^0\widetilde q_{kL}\widetilde q_{kL}
\qquad\qquad&{ig^2\sin2\alpha\over 2}\left[{T_{3k}-e_k\sin^2\theta_W
\over\cos^2\theta_W}-{m_q^2\over2m_W^2}D_k\right]\,,\crr
H_1^0H_2^0 \widetilde q_{kR}\widetilde q_{kR}
\qquad\qquad&{ig^2\sin2\alpha\over 2}\left[e_k\tan^2\theta_W
-{m_q^2\over2m_W^2}D_k\right]\,.\cr}$$
\point
In eq.~(4.27), in the term proportional to $N\psi_N\psi_N$ remove the
factor of 2.
\point
In eq.~(4.28), add $+ {\rm h.c.}$ to the right hand side.
\point
Eq. (4.29) should read ${\cal L}_m^{soft}=\half M \lambda^a\lambda^a
+\half M^\prime \lambda^\prime\lambda^\prime+{\rm h.c.}$.
\point
In the second line of eq.~(4.30), replace $v_2$ with $-v_2$; \ie\
$(v_1\psi_{H_1}^0-v_2\psi_{H_2}^0)$ is the correct form.
\point
In eq.~(4.48), replace $+\sqrt2$ with $\pm\sqrt2$.
In eq.~(4.49), replace $-\sqrt2$ with $\mp\sqrt2$.
In both cases, the upper sign should be used for $H_1^0$ and $H_2^0$
couplings to $\wtil\chi^0\wtil\chi^0$ given in eq.~(4.47), and
the lower sign should be used for the corresponding couplings of
$H_3^0$.
\point
In Fig. 22, replace $(1-\gamma_5)$ by $(1+\gamma_5)$ in (b) and (d).
In Fig. 23, replace $(1+\gamma_5)$ by $(1-\gamma_5)$ in (b) and (d).
\point
In the second line of eq.~(5.5), the plus sign immediately following $P_L$
should be a minus sign.
Note that the corresponding Feynman rules depicted in Fig. 24 are correct.
\point
In eq.~(A.2), insert: $+ {\rm h.c.}$ at the end.

\endpage

\bigskip
\titlestyle{Errata for J.F. Gunion and H.E. Haber,}
\titlestyle{\npbj{278} (1986) 449.}
\medskip
\pointbegin
In eq.~(A.16), replace $\eps$ with $-\eps$.

\bigskip
\titlestyle{Errata for J.F. Gunion and H.E. Haber,}
\titlestyle{\npbj{307} (1988) 445.}
\medskip
\pointbegin
In eq.~(A.7), replace $\left(1-\mz^2/m_H^2\right)^{1/2}$ by
$\left(1-4\mz^2/m_H^2\right)^{1/2}$.
\point
Eqs.~(A.8) and (A.9) are more conveniently written as:
$$\Gamma(\hp\rta W^+ H_2^0)={g^2\lam^{3/2}\cos^2(\beta-\alpha) \over
               64\pi\mw^2\mhp^3},\eqno{(A.8)}$$
$$\Gamma(H_3^0\rta Z^0 H_2^0)={g^2\lam^{3/2}\cos^2(\beta-\alpha) \over
               64\pi\mz^2m_{H_3^0}^3 \cos^2\theta_W},\eqno{(A.9)}$$
\bye